# Entropy-Guided Greedy Heuristic for Minimal Differential Set Construction

Mariana del Fresno[1,2], María del Carmen Romero[1], Alejandro Clausse[3,*]

[1]Universidad Nacional del Centro de la Provincia de Buenos Aires, 7000 Tandil, Argentina.
[2]Comisión de Investigaciones Científicas de la Provincia de Buenos Aires, 1900 La Plata, Argentina.
[3]Instituto Tecnológico de Buenos Aires, Iguazú 341, 1437 CABA, Argentina.
[*]Corresponding author: clausse@exa.unicen.edu.ar

## Abstract

This paper addresses the challenge of identifying a minimal subset of discrete predictor variables that best explains a binary class. We propose a greedy heuristic for feature selection that sequentially constructs a differential set by iteratively adding the variable that yields the most significant reduction in conditional entropy. The method incorporates a confidence-guided acceptance criterion that accounts for finite-sample uncertainty, allowing the search to adaptively balance the risk of including spurious variables against the opportunity to capture meaningful predictors. Because the procedure yields a compact set of statistically validated predictors together with an explicit confidence level for each one, it is well suited to healthcare innovation and management applications such as biomarker and risk-factor identification, diagnostic panel design, and clinical decision support, where minimizing the number of tests or indicators collected is critical for cost, feasibility, and patient burden.

## 1. Introduction

The identification of a minimal subset of features that optimally predicts a target variable is a cornerstone of machine learning and data science, with critical applications in bioinformatics, clinical diagnostics, healthcare management, finance, and pattern recognition (Saeys et al., 2007). Feature selection improves model interpretability, enhances generalization, and reduces computational cost by removing irrelevant or redundant variables (Guyon and Elisseeff, 2003).

Filter methods based on mutual information remain popular due to their simplicity and efficiency. For instance, the widely used Minimum Redundancy Maximum Relevance criterion (Peng et al., 2005) performs incremental selection, greedily adding features that maximize relevance to the target while minimizing redundancy with already-selected features. While effective, such methods often rely on point estimates of information-theoretic quantities and can be sensitive to finite-sample estimation errors, an issue particularly acute for non-normal distributions, where convergence of entropy estimators can be slow and confidence intervals difficult to calibrate (Grigorenko and Grigorenko, 2024).

Heuristic approaches offer computationally tractable alternatives to exhaustive search, which is particularly valuable when the variable space is large (Chandrashekar and Sahin, 2014). Recent reviews, such as Faizan and Mohammad Arif (2025), highlight the widespread use of swarm intelligence and evolutionary algorithms, including genetic algorithms and particle swarm optimization, for identifying optimal feature subsets. Information-theoretic criteria such as entropy have likewise been integrated into metaheuristic optimization frameworks in other domains, underscoring the versatility of entropy as an objective function within heuristic search.

Among the many ways to frame this problem, the information-theoretic approach seeks the subset of variables carrying the most information about the target, quantified via mutual information or conditional entropy (Cover and Thomas, 2006; see also Mielniczuk, 2022). The ideal solution selects the subset $D$ minimizing the conditional entropy of the target $C$ given the features, $H(C/D)$; this is a combinatorial problem, and exhaustive evaluation of all subsets is NP-complete.

To address estimation uncertainty, several lines of inquiry have emerged, including Bayesian modeling of the posterior distribution of mutual information (Archer et al., 2014) and bootstrap-based techniques to assess feature-set stability (Nogueira et al., 2017). There is also growing recognition of the need for methods that provide statistical confidence in selection decisions, moving beyond purely greedy optimization based on point estimates. Chen et al. (2022) provide practical guidelines for evaluating conditional entropy and mutual information, emphasizing robust estimation in discovering major factors.

Concurrently, the theoretical understanding of entropy minimization itself is evolving. Press et al. (2024) explore the so-called entropy enigma, analyzing when entropy minimization succeeds or fails, underscoring the need for careful, statistically grounded application. The challenge of reliable estimation with limited data has also motivated dynamic feature selection methods that learn to maximize mutual information (Covert et al., 2023). However, sequential selection procedures that explicitly incorporate finite-sample uncertainty of entropy estimates into the decision rule itself have received little attention, particularly for identifying a minimal differential set. In this work, we address this gap by proposing a greedy heuristic for differential set selection, grounded in information channel theory, that explicitly incorporates finite-sample uncertainty into the selection rule. Unlike conventional greedy algorithms relying on point estimates, our approach derives confidence bounds for entropy estimates, extending Grigorenko and Grigorenko (2024), and uses them as an adaptive acceptance criterion that governs both variable selection and search termination, balancing exploration of uncertain variables against exploitation of those with clear entropic benefits, while maintaining computational efficiency through its greedy construction. The method offers an alternative in low-data regimes.

Beyond its general relevance to machine learning, constructing a minimal, statistically defensible predictor set is especially pressing in healthcare innovation and management. Clinical decision-making increasingly relies on panels of biomarkers, risk factors, or diagnostic indicators,for instance to predict long-term disease risk from patient variables (Ahmed and Shefaq, 2022) or to extract informative features from electronic health records (Lu and Segall, 2024), yet collecting and processing each additional variable carries direct costs in laboratory resources, patient burden, and clinician time (Remeseiro and Bolón-Canedo, 2019). A method able to identify a small, high-confidence set of explanatory variables from limited samples, as is common in clinical studies and pilot cohorts, can inform leaner diagnostic protocols and evidence-based resource-allocation policies. The confidence-guided acceptance criterion introduced here supports this goal by attaching an explicit statistical confidence level to each candidate predictor, giving healthcare managers and clinicians a transparent basis for deciding which indicators are worth the cost of routine collection.

## 2. Problem formulation

The problem of interest is the dependence of a binary class $C \in \{0,1\}$ (e.g., sick/healthy), on a set $V = \{X_1, X_2, X_3, \dots X_{N_X}\}$ of stochastic discrete variables that are mutually independent. A finite set of $m$ sample observations of the values of $V$ and $C$ is available. The primary goal is to identify the subset of variables that are significant predictors of the class, given the observed data. We will say that this subset of variables "explains" $C$, and it is referred to as the *differential set*.

The forementioned problem can be treated using information channel theory. An information channel represents the relationship between two sources of information, one input and one output. In our case, the output source is the class $C$, and the input source is composed of the $N_X$ independent variables $X_i$. Within that framework, the statistical indicator for evaluating the degree of significance of a subset of variables $D \subset V$ is the conditional entropy (a.k.a. noise), defined as:

$$H(C|D) = \sum_{D} p(D) h(q_D) \tag{1}$$

where the summation is taken over all possible realizations of the subset $D$, $p(D)$ denotes the probability of each such realization, and $q_D$ is the conditional probability that the class $C$ takes the value 0 given that the subset takes the value $D$:

$$q_D = Prob(C = 0|D) \tag{2}$$

The function $h(x)$ represents the Shannon entropy of a binary variable with probability $x$, that is:

$$h(x) = -x \log x - (1 - x) \log(1 - x) \tag{3}$$

Using base 2 for the logarithm, $H(C|D)$ quantifies the average number of binary questions (i.e., questions whose answer can only be yes or no) that would need to be asked to fully explain the class $C$, given that the values of the variables of $D$ are known. If the subset $D$ completely determines the class, then $H(C|D) = 0$; conversely, if the class is independent of $D$, then $H(C|D) = H(C)$ Therefore, the problem consists in identifying the subset $D$ that minimizes $H(C|D)$.

To estimate $H(C|D)$ based on the observation sample, it is necessary to first estimate the joint probability distribution $p\left(C, X_1,\ X_2, X_3, \dots, X_{N_X}\right)$ and then compute $H(C|D)$ for all possible subsets. Subsequently, we determine the differential set according to an optimization criterion, for example minimizing $H(C|D)$. This problem is NP-complete, and thus becomes computationally expensive when the number of variables is large.

The method proposed in this work is an estimation of the differential set via a selective pruning procedure, which significantly reduces computational cost. Starting from the estimated joint distribution $p\left(C, X_1, X_2, X_3, \dots, X_{N_X}\right)$, the method constructs a sequence of differential subsets $D_n$ , $n = 1, 2, 3, \dots$, where at each iteration $n$ the variable that most reduces $H(C|D_n)$ is added to the subset.

In theory, variables that are irrelevant to the class do not contribute any information; therefore, the noise remains unchanged when any such variable is added to $D$. That is, if a variable $X_j$ is irrelevant to the class, then:

$$H\left(C \middle| D_n X_j\right) = H(C|D_n) \tag{4}$$

Accordingly, it is expected that $H(C|D_n)$ will decrease with $n$ until reaching a minimum value, beyond which the inclusion of additional variables no longer contributes to reducing the noise.

In practice, however, finite-sample estimation introduces uncertainty that prevents direct verification of Eq. 4, motivating the confidence-guided heuristic approach developed in the next section.

## 3. **General Procedure for Constructing the Sequence of Differential Subsets**

Given the finite sample size and the inherent uncertainty in entropy estimation, a purely deterministic greedy selection would be susceptible to spurious selections. To address this, we frame the selection process as a heuristic search where the decision to include a variable is governed by an adaptive confidence threshold. This threshold controls the trade-off between aggressively seeking entropy reduction and conservatively avoiding overfitting. At each iteration, the algorithm evaluates candidate variables using a heuristic scoring function that combines the estimated entropy reduction with its associated uncertainty, which will be expressed through a single parameter $k$.

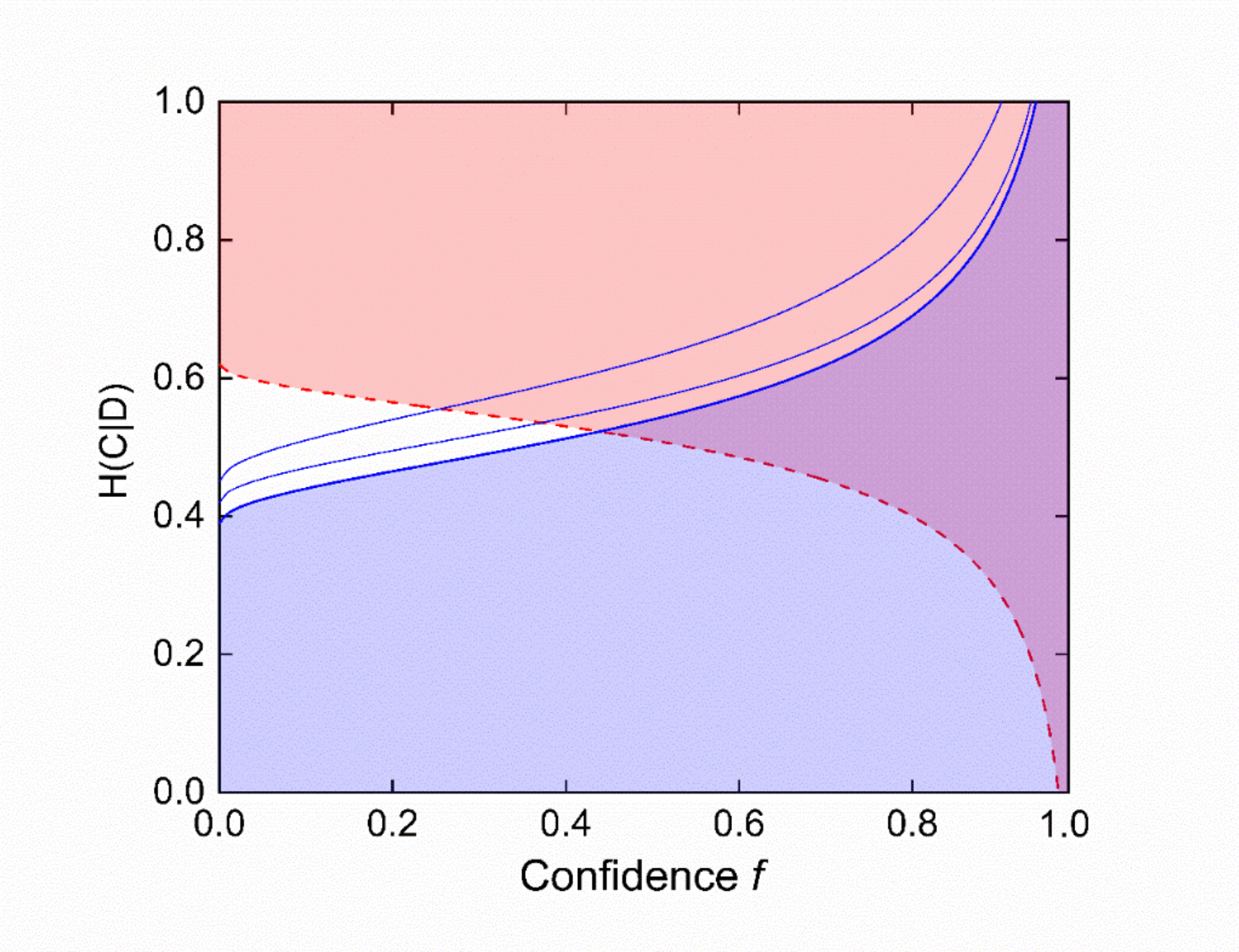

Figure 1. Lower (dashed red) bound of the conditional entropy of the differential subset $D_n$ and upper bounds of the conditional entropy of the differential subset $D_{n+1}$.

Let us assume that given a finite set $N_{obs}$ of sample $m$ observations of the values of the subset $V$ and $C$, we can produce an estimate of the conditional entropy $H_{est}$ of the conditional entropy $H_{real}$ and its standard deviation $\sigma_{est}$. Given these estimates, there exist several theorems providing upper and lower bounds for the estimation, which guarantee a given confidence level $f$ (Lin and Bai, 2010). Several of these bounds generally take the form:

$$P(H_{real} \geq H_{est} - k\,\sigma_{est}) = P(H_{real} \leq H_{est} + k\,\sigma_{est}) \geq f(k^2) \qquad (5)$$

We will rely on Cantelli's criterion (Cantelli, 1910), according to which:

$$f(k^2) = \frac{k^2}{1 + k^2} \qquad (6)$$

The intersection between the lower probability bound of the differential subset $D_n$ and the upper bound of each possible subset $D_{n+1}$ provides a confidence indicator for each of the latter.

The search procedure is illustrated in Figure 1, which visualizes the heuristic decision boundary used to select variables. The red-shaded area represents the current solution's performance baseline (i.e., the conditional entropy of the current differential set) with its associated uncertainty captured by the lower confidence bound, to wit:

$$P\left(H_{n,real} \le H_{n,est} + \sqrt{\frac{f}{1-f}}\,\sigma_{n,est}\right) \ge f \tag{7}$$

where $k$ was expressed in terms of the confidence level $f$ by means of Eq. 6. The blue curves represent the candidate solutions formed by adding each remaining variable, that is:

$$P\left(H_{n+1,real} \ge H_{n+1,est} - \sqrt{\frac{f}{1-f}}\,\sigma_{n+1,est}\right) \ge f \tag{8}$$

The intersection points between these bounds provide a scoring mechanism (the parameter $f$) that ranks candidate variables by the statistical significance of their contribution, analogous to how heuristic methods use objective functions to compare candidate solutions.

The intersection of the lower blue curve with the red curve provides an estimator of the maximum confidence that we can guarantee for incorporating the new variable into $D$. For values of $f$ greater than the intersection the entropy bands overlap and we can no longer ensure that the new variable contributes significantly to reducing the conditional entropy of the class. Since Cantelli's function guarantees a strictly increasing bijective relationship between $f$ and $k$, the intersection can be worked out in terms of $k$. For instance, once the first differential subset $D_1$ has been selected, the lower bound of the conditional entropy of the class is given by:

$$H_{inf,1} = H_{est,1} - k\,\sigma_{est,1} \tag{9}$$

For each variable $X_j \notin D_1$, a possible extended subset is constructed as $D_{2,j} = D_1 \cup \{X_j\}$, whose upper entropy bound is given by:

$$H_{sup,j} = H_{est,j} + k\,\sigma_{est,j} \tag{10}$$

Both bounds intersect at the point where $H_{inf,1} = H_{sup,j}$. Solving for $k_j$ we obtain:

$$k_j = \frac{H_{est,1} - H_{est,j}}{\sigma_{est,1} + \sigma_{est,j}} \tag{11}$$

The variable $X_j \notin D_1$ that maximizes $k_j$ is selected, as it ensures the entropic contribution to $D_2$ with higher confidence. This new variable is then incorporated into the differential subset $D_2$.

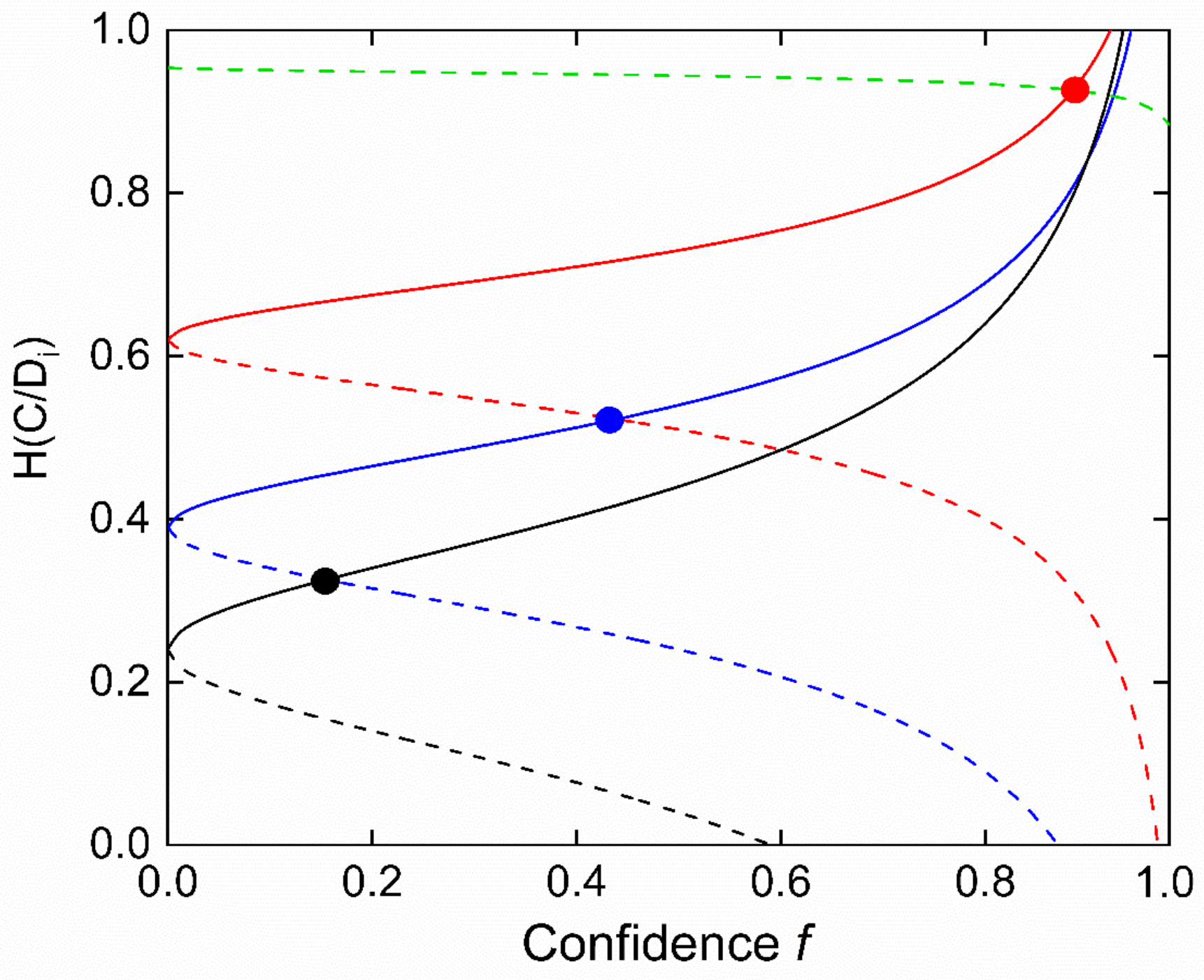

Figure 2. Illustration of the selection algorithm applied to the differential set. The first three variables selected are highlighted in red, blue, and black, corresponding to their relative influence on the class discriminant. The green line is the estimation of the lower bound of the unconditional entropy of the class. The iterative selection process is defined by the intersection points, indicated with circles, between the solid upper bounds and the dashed lower bounds of the corresponding sequent variables.

The confidence $f(k_1)$ associated with the first subset $D_1$ can be estimated by calculating the value of $k$ for which the upper bound of entropy reaches the unconditional entropy of the class, $H(C)$. That is:

$$H_{est,1} + k_1\, \sigma_{est,1} = H_{est}(C) + k_1\, \sigma_{est}$$

$$k_1 = \frac{H_{est}(C) - H_{est,1}}{\sigma_{est,1} + \sigma_{est}} \quad (12)$$

Figure 2 illustrates the first three iterations of a sequential construction of the differential set. At each step, a new variable is selected and included in the set based on the intersection points from the previous step.

The general procedure for constructing the sequence of differential subsets is then the following:

1) *Initial Estimation*

   Estimate the entropy of the class $H(C)$ and the conditional entropies $H_i^{(1)} = H(C|X_i)$ for all variables $X_i$, together with their corresponding standard deviations $\sigma_{est}$ and $\sigma_i^{(1)}$. Identify the variable $X^{(1)}$ that maximizes:

$$k_{1,i} = \frac{H(C) - H_i^{(1)}}{\sigma_i^{(1)} + \sigma_{est}} \tag{13}$$

2) *First Differential Subset*

   Define the first differential subset as:

$$D_1 = \{X^{(1)}\} \tag{14}$$

   The contribution of $X^{(1)}$ to the explanation of the class corresponds to a reduction in conditional entropy by an amount $\Delta H^{(1)} \geq k^{(1)}\sigma^{(1)}$ with confidence $f(k^{(1)})$, where $k^{(1)} = \max k_{1,i}$.

3) *Iterative Expansion*

   Estimate $H_i^{(2)} = H(C|X_i)$ and its corresponding standard deviation $\sigma_i^{(2)}$ for each variable $X_i \notin D_1$. Identify the variable $X^{(2)}$ that maximizes:

$$k_{2,i} = \frac{H^{(1)} - H_i^{(2)}}{\sigma^{(1)} + \sigma_i^{(2)}} \tag{15}$$

   Define the second differential subset as:

$$D_2 = D_1 \cup \{X^{(2)}\} \tag{16}$$

   The contribution of $X^{(2)}$ to the explanation of the class corresponds to a reduction in conditional entropy by an amount $\Delta H^{(2)} \geq k^{(2)}(\sigma^{(1)} + \sigma^{(2)})$ with confidence $f(k^{(2)})$, where $k^{(2)} = \max k_{2,i}$.

4) *Termination Criteria*

   The termination criteria are designed to reflect the convergence conditions commonly found in heuristic search algorithms. Specifically, the search stops when one of the following conditions is met:

- Exhaustion of search space: no remaining variables to evaluate.

- Confidence threshold: the confidence level $f$ of the differential subset falls below a user-defined minimum, analogous to a cooling schedule in simulated annealing.
- Stagnation: no positive $k$ values are found, indicating that that no further improvement is detected.

Appendix 1 describes the details of a sample's estimation of the conditional entropy and its standard deviation.

## 4. Case Study

To test the method, we define a test case with a binary class $C \in \{0,1\}$ and five mutually independent, uniformly distributed binary features $\{X_1, X_2, X_3, X_4, X_5\}$. Binary variables were chosen for simplicity, but the method is applicable to variables with more than two states. The class label $C$ is conditionally dependent only on $X_1$ and $X_2$, with their relationship specified by the distribution $P(X_1, X_2, C)$ in Table 4. Therefore, the overall joint distribution factors as:

$$P(X_1, X_2, X_3, X_4, X_5, C) = \frac{P(X_1, X_2, C)}{8} \quad (17)$$

The joint distribution yields the following theoretical conditional entropies:

$H(C|X_1) = 0.6661$

$H(C|X_2) = 0.8425$

$H(C|X_3) = H(C|X_4) = H(C|X_5) = 0.9544$

Table 4. Joint probability distribution $P(X_1, X_2, C)$

| $x_1$ | $x_2$ | $c$ | $P(X_1, X_2, C)$ |
|---|---|---|---|
| 0 | 0 | 0 | 23/96 |
| 0 | 0 | 1 | 1/96 |
| 0 | 1 | 0 | 3/32 |
| 0 | 1 | 1 | 5/32 |
| 1 | 0 | 0 | 1/24 |
| 1 | 0 | 1 | 5/24 |
| 1 | 1 | 0 | 0 |
| 1 | 1 | 1 | 1/4 |

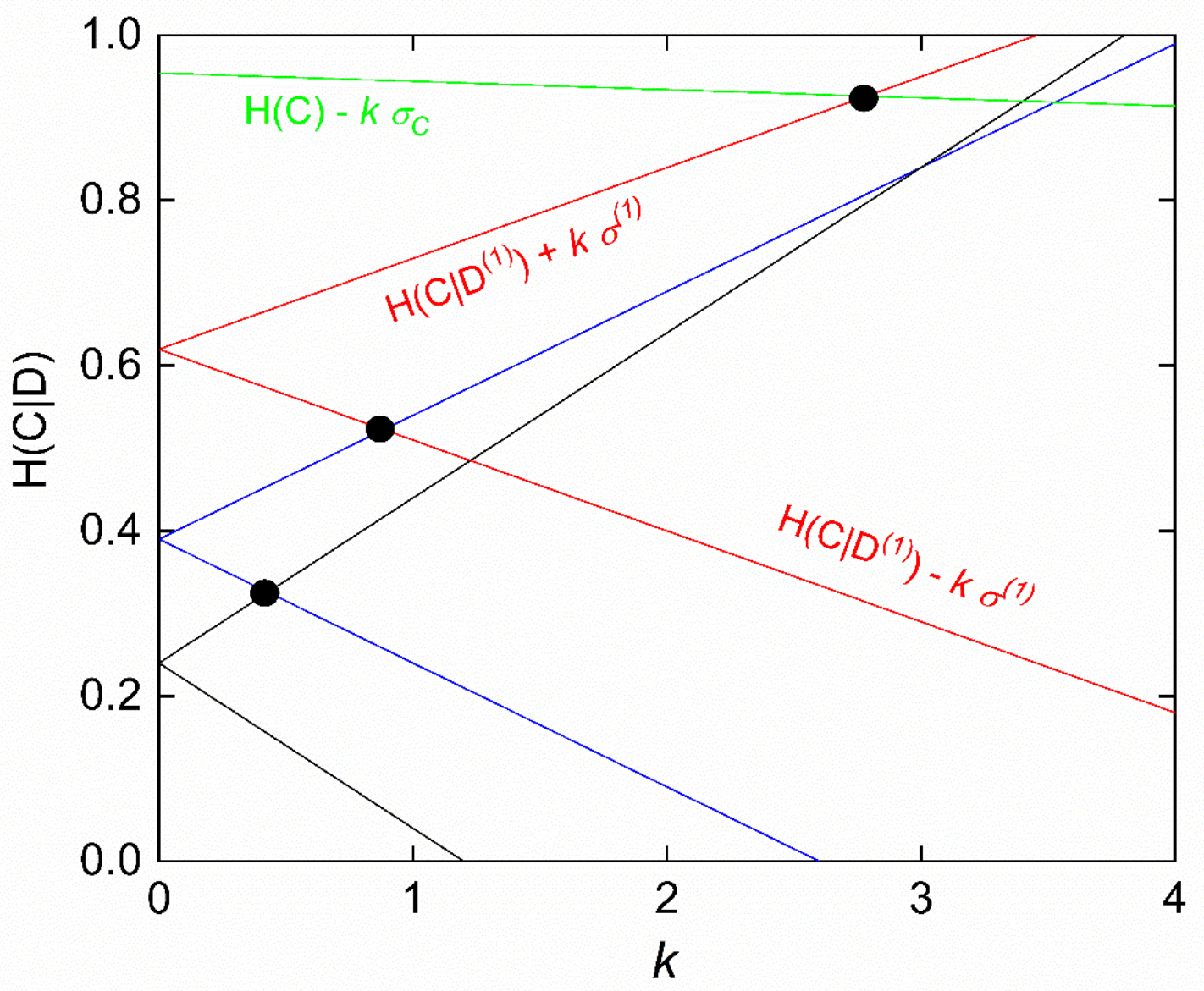

Figure 3. Illustration of the selection algorithm applied to the differential set interms of the equivalent parameter $k$. The first three variables selected are highlighted in red, blue, and black, corresponding to their relative influence on the class discriminant. The green line is the estimation of the lower bound of the unconditional entropy of the class. The iterative selection process is defined by the intersection points, marked with black circles.

We used the joint distribution $P(X_1, X_2, X_3, X_4, X_5, C)$ to generate 10,000 simulated datasets for each sample size $m = 10$, 20 and 50 observations each. This design models real-world data collection with limited samples. We then applied the method from Section 3 to each dataset to evaluate its performance under these constraints. Figure 3 illustrates the first three iterations of the sequential construction of the differential set using the values obtained in the case study. At each step, the algorithm selects a new variable (color-coded by class influence) based on the intersection points (black circles) from the previous step.

Table 5 presents the frequency with which each variable was selected in each iteration, across different sample sizes. As expected, in the first iteration, the variable selected most frequently was $X_1$ (shown in red); with a sample size of 50, it was selected in 87.8% of the simulations. Performance declined with smaller sample sizes of 10 and 20 observations, as anticipated, due to the reduced information available in each dataset. The variable $X_2$ (shown in blue), which has a weaker influence on the class, was selected in 12% of the simulations with a sample size of 50. This percentage increased up to 24% with a sample size of 10. The remaining variables $(X_3, X_4, X_5)$ have no true influence on the class, which is reflected in the minimal rate of spurious selections across all conditions. The accuracy of the method (percentages of true outcomes) corresponds to the red values of the first iteration and the blue values in the second iteration.

Figure 4 shows box plots of the conditional entropy $H(C|D^{(n)})$ of the class $C$, obtained by testing each remaining variable's addition to the differential set at iteration $n$. Results are grouped by sample size and iteration (1–5, top to bottom); the most influential variable, $X_1$, is shown in red, the second, $X_2$, in blue, and the remaining non-influential variables, $X_3$, $X_4$, $X_5$, in grey. As expected, smaller samples widen the distributions, reflecting greater uncertainty.

Table 5. Selection frequency (%) per variable in each iteration for different sample sizes

| Iteration | Sample size | | | Variable |
|---|---|---|---|---|
| | 10 | 20 | 50 | |
| 1 | 59.13 | 73.57 | 87.80 | $X_1$ |
| | 24.23 | 22.01 | 12.15 | $X_2$ |
| | 5.64 | 1.53 | 0.00 | $X_3$ |
| | 5.61 | 1.38 | 0.04 | $X_4$ |
| | 5.39 | 1.51 | 0.01 | $X_5$ |
| 2 | 22.08 | 20.35 | 12.18 | $X_1$ |
| | 31.72 | 52.76 | 85.00 | $X_2$ |
| | 15.43 | 9.01 | 0.83 | $X_3$ |
| | 15.88 | 8.84 | 1.03 | $X_4$ |
| | 14.90 | 9.04 | 0.96 | $X_5$ |
| 3 | 10.80 | 3.17 | 0.01 | $X_1$ |
| | 19.28 | 11.42 | 1.95 | $X_2$ |
| | 25.50 | 28.86 | 32.78 | $X_3$ |
| | 23.39 | 27.71 | 32.18 | $X_4$ |
| | 21.02 | 28.84 | 33.07 | $X_5$ |
| 4 | 9.81 | 1.76 | 0.01 | $X_1$ |
| | 16.92 | 7.05 | 0.48 | $X_2$ |
| | 25.57 | 30.81 | 32.90 | $X_3$ |
| | 24.38 | 31.14 | 33.62 | $X_4$ |
| | 23.32 | 29.24 | 32.99 | $X_5$ |
| 5 | 7.00 | 1.81 | 0.00 | $X_1$ |
| | 19.00 | 8.24 | 0.42 | $X_2$ |
| | 27.33 | 30.12 | 33.40 | $X_3$ |
| | 26.33 | 30.01 | 33.19 | $X_4$ |
| | 20.33 | 29.82 | 32.99 | $X_5$ |

Table 6. Percentage of stops per iteration. A minimum value of $k = 0.01$ was imposed.

| Iter | Observations | | |
|---|---|---|---|
| | 10 | 20 | 50 |
| 1 | 13.3 | 0.9 | 0 |
| 2 | 41.3 | 5.6 | 0.04 |
| 3 | 77.4 | 20.0 | 0.1 |
| 4 | 97 | 63.0 | 2.0 |

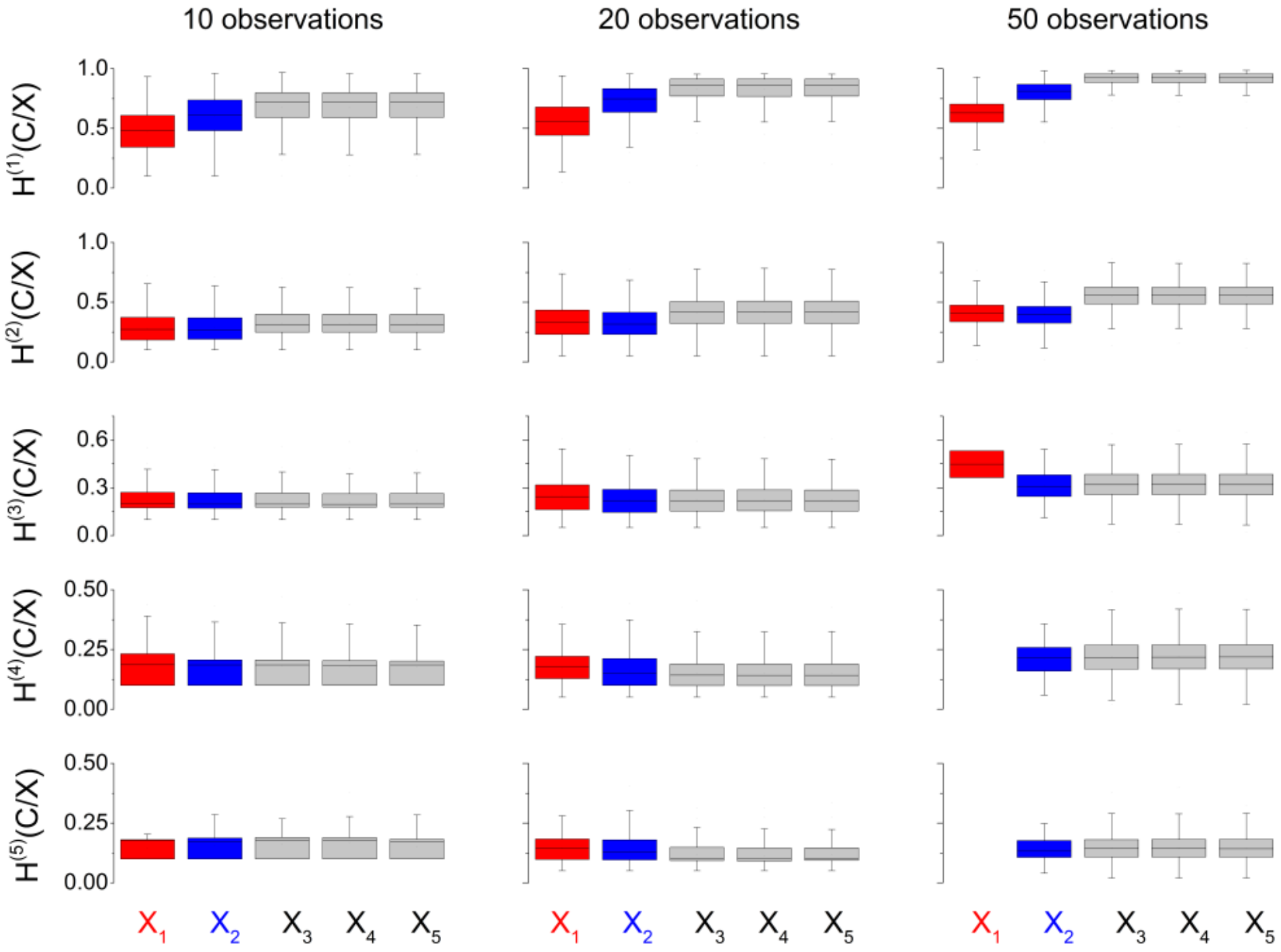

Figure 4. Box plots of conditional entropy values obtained by sequentially adding each variable to the differential set. Results are grouped by sample size (10, 20, and 50) and displayed by iteration (from 1 to 5, top to bottom). The most influential variable in each case is highlighted in red, the second most influential in blue, and all remaining (non-influential) variables in grey.

Consider the first iteration (top row of Figure 4), analyzing the class's dependence on each variable individually. At n=50, the entropy distribution for $X_1$ lies clearly below that for $X_2$, though with partial overlap, explaining why $X_2$ was occasionally selected; this overlap grows more pronounced at n=20 and n=10. In the second iteration, entropy values shift lower, reflecting a two-variable differential set, and the boxes for $X_1$ and $X_2$ share a similar mean entropy, consistent with both variables typically appearing together.

Figure 5 shows box plots of the selection parameter $k$ for each remaining variable at iteration $n$, using the same layout and colors as Figure 4. In the first iteration, $X_1$ reaches the highest $k$ values (mean 1.17 at n=50, versus 0.65 for $X_2$). At n=10 and n=20, negative $k$ values occasionally occur, indicating no discernible influence, consistent with the lower survival rate (Table 6) at smaller sample sizes.

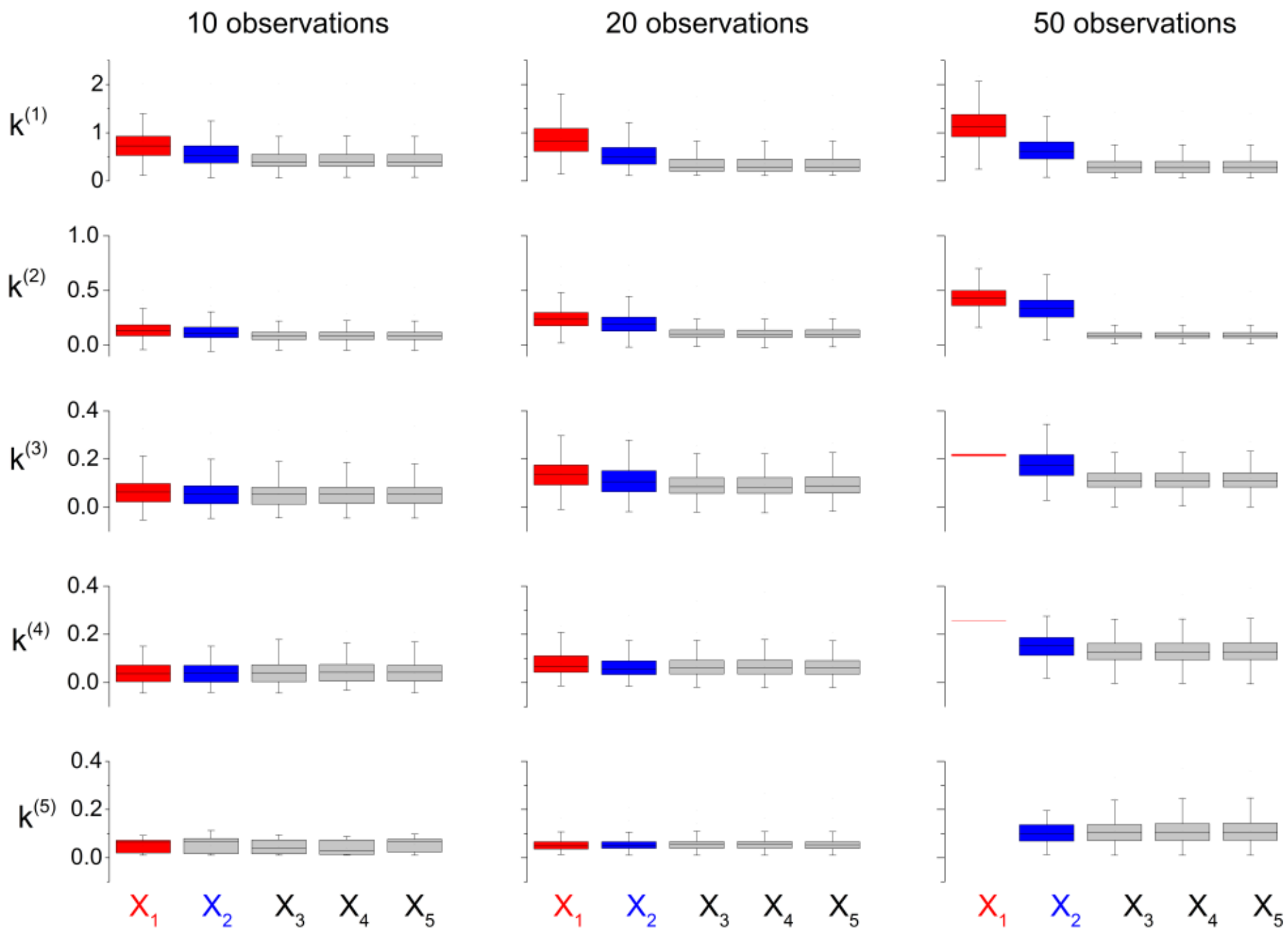

Figure 5. Box plots of the selection parameter $k$ obtained by sequentially adding each variable to the differential set. Results are grouped by sample size (10, 20, and 50) and displayed by iteration (from 1 to 5, top to bottom). The most influential variable in each case is highlighted in red, the second most influential in blue, and all remaining (non-influential) variables in grey.

The most informative indicator is the proportion of trials in which the method exactly recovers the true differential set $\{X_1, X_2\}$, i.e., $X_1$ selected first, $X_2$ second, with no further variables added. Fig. 6 plots this indicator against the cutoff value $k$ over 1000 trials, for sample sizes of 10, 20, and 50 (solid curves: 3 irrelevant variables; dashed: 8). Sample size has the strongest effect; the number of irrelevant variables plays only a secondary role. Low cutoffs admit irrelevant variables alongside $X_1$ and $X_2$; high cutoffs cause premature termination, typically missing the weaker true variable, the classic exploration–exploitation trade-off of heuristic search.

Table 7. Mean and standard deviations of $H(C|D^n)$ and $k$ obtained in the first three iterations.

| Iteration | 10 observations | | 20 observations | | 50 observations | |
|---|---|---|---|---|---|---|
| | ***H*** | ***k*** | ***H*** | ***k*** | ***H*** | ***k*** |
| **1** | 0.48 ± 0.20 | 0.87 ± 0.35 | 0.56 ± 0.17 | 0.88 ± 0.39 | 0.62 ± 0.11 | 1.17 ± 0.37 |
| **2** | 0.28 ± 0.12 | 0.12 ± 0.08 | 0.32 ± 0.13 | 0.20 ± 0.11 | 0.39 ± 0.10 | 0.34 ± 0.11 |
| **3** | 0.21 ± 0.08 | 0.06 ± 0.05 | 0.22 ± 0.09 | 0.09 ± 0.05 | 0.32 ± 0.09 | 0.12 ± 0.05 |

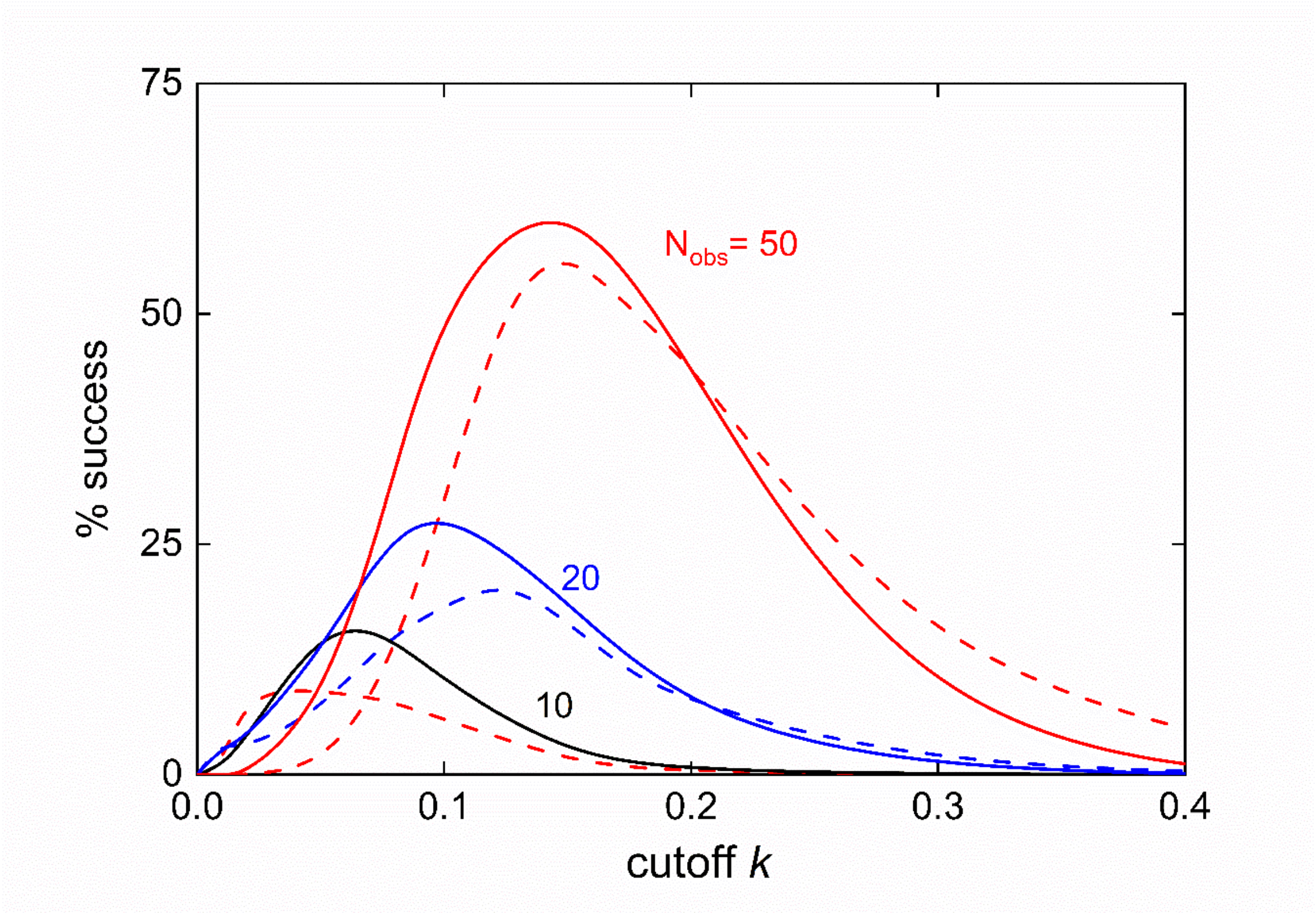

Figure 6. Proportion of trials in which the method correctly identifies the true differential set $\{X_1, X_2\}$ and terminates without selecting additional variables. Solid curves represent the scenario with three irrelevant variables in addition to $\{X_1, X_2\}$, whereas dashed curves correspond to the scenario with eight irrelevant variables.

To contextualize our method, we compared its performance against two established approaches: VarImp (Kuhn, 2021), a filtering criterion, and SES (Tsagris et al., 2019), which identifies statistically equivalent signatures. Table 8 shows the confidence-guided heuristic consistently outperforming both alternatives across all sample sizes; its ability to report a confidence level for each selected variable (average lower bounds of 0.46–0.65 for $X_1$ and 0.10-0.23 for $X_2$) offers decision-support information beyond conventional feature-selection methods.

Table 8. Percentage of trials (over a total of 1000 trials) in which each method selected the differential set $\{X_1, X_2\}$. For the proposed method, only the results from the first and second iterations were considered. For VarImp, the two highest-ranked variables were taken into account. For SES, only differential sets consisting of two variables were included.

| Sample size | Present method | VarImp | SES |
|---|---|---|---|
| 30 | 77.55 | 77.24 | 76.43 |
| 50 | 88.65 | 78.40 | 78.68 |
| 100 | 97.06 | 82.75 | 81.23 |

## Conclusions

We presented a greedy heuristic for identifying a minimal differential set of variables that significantly predicts a binary class, framing the problem within information channel theory and adopting conditional entropy minimization as the optimization objective. Its main contribution is a confidence-guided acceptance criterion that explicitly incorporates the statistical uncertainty of entropy estimation from finite samples, adaptively governing both variable selection and termination while balancing exploration and exploitation under uncertainty.

At each step, the procedure selects the variable yielding the largest statistically significant reduction in conditional entropy, as determined by a heuristic scoring function (the parameter $k$). This constructive approach offers a computationally efficient alternative to exhaustive search, with the confidence-guided rule providing robustness against spurious selections and convergence via stagnation, confidence threshold, or search-space exhaustion, without requiring problem-specific parameter tuning.

A simulation study with a known true differential set validated the method's effectiveness: it correctly identified the most influential variables, with accuracy decreasing as sample size decreased, as expected, while the rate of spurious selection of irrelevant variables remained low across all tested sample sizes.

From a practical, healthcare-management standpoint, the method supports evidence-based decisions about which clinical or operational indicators are worth collecting: since it halts once further variables no longer provide statistically credible entropy reduction, it discourages retaining costly but uninformative tests, while its confidence levels give managers a quantitative basis for judging a new variable's added value.

The present formulation assumes independence among predictor variables, simplifying entropy estimation and the sequential selection procedure. This is a limitation, but it does not preclude applying the method to interdependent predictors: variables excluded from the differential set can be interpreted as irrelevant, or as having their relevance already captured by the selected variables' dependence structure, so the method remains useful by highlighting the most significant contributors while implicitly accounting for redundancy.

In summary, the proposed method offers an information-theoretically grounded, computationally tractable framework for feature selection under finite-sample uncertainty. Future work will apply it to real-world clinical and biomedical datasets for biomarker and risk-factor selection in disease diagnosis and healthcare resource management, and explore alternative confidence bounds to further improve its statistical power.

**Consent to Participate declaration:** not applicable.

**Consent to Publish declaration:** not applicable.

**Ethics declaration:** not applicable.

**Conflict of interest statement**: The authors confirm that there is no conflict of interest related to the article.

**Data availability**: The data that support the findings of this study are available from the corresponding author upon reasonable request.

## Appendix 1. Estimation of the conditional entropy and its standard deviation

The direct method to estimate the entropy of a finite binary sample of size $m$ is by counting the frequency of one of the binary states and then calculating the Shannon function given by Eq. 3. However, this estimate has a systematic bias that should be corrected by adding the Miller correction $(2\ln 2\ m)^{-1}$ (Hausser and Strimmer, 2014). To estimate the standard deviation of the entropy estimator, the sample is partitioned into smaller subsets using a jackknife approach. For each subset $j$, an entropy estimate $H_{j,est}$ is computed. The overall entropy estimate $H_{est}$ is then obtained as the average of these values, while the associated standard deviation $\sigma_{est}$ is derived from their dispersion. In this work, we used subsets of size $m/2$. To mitigate spurious cases in which small sample sizes yield a vanishing standard deviation, we introduce a regularization correction term, $(m/2)^{-1/2}$, to $\sigma_{est}$. This adjustment prevents division by zero in Eqs. (11) and (12).

Since the theoretical number of possible partitions is combinatorial, a practical cap can be imposed to control computational cost (e.g., 5000 subsets). Thus, for small samples all possible halves are considered, whereas for larger samples 5000 random partitions are selected following a Monte Carlo resampling scheme. This strategy mitigates finite-sample bias and reduces sensitivity to distributional irregularities, yielding a corrected standard deviation estimate that more reliably reflects the uncertainty of entropy measurements in practice. Following is a pseudocode of the described procedure:

**Input:**

$N_{obs}$: sample with *m* observations of the dataset containing $\{C, X_1,\ X_2, X_3, \dots X_{N_X}\}$

where $C \in \{0,1\}$ is the binary class and $X_1,\ X_2, X_3, \dots X_{N_X}$ the stochastic variables

D: current differential set of selected variables

**Output:**

*H*: estimated conditional entropy $H(C|D)$

$\sigma_H$: standard deviation of $H(C|D)$ across sub-samples

1. *Initialization*

$m$ ← number of observations in $N_{obs}$

$n$ ← ⌊ m / 2 ⌋ # size of each sub-sample from $N_{obs}$

$N_{sub}$ ← number of random sub-samples of size $n$

$H_{sub}$ [1, ..., $N_{sub}$] ← 0 # H per sub-sample to store entropy estimates

2. *Sub-sampling and entropy estimation*

For s = 1 to $N_{sub}$

Randomly select $n$ rows from $N_{obs}$ (class $C$ and variables in $D$) → ( $C_{sub}$, $D_{sub}$)

For each unique combination $d_j$ in $D_{sub}$ :

$p(D = d_j) \leftarrow N_j / n$ # proportion of samples with D = $d_j$

$q_{d_j} \leftarrow p(C = 0 \mid D = d_j)$ # proportion of class−0 samples within $d_j$

$H(C \mid D = d_j) \leftarrow entropy_bin(q_{d_j})$ # function entropy with Miller correction.

$H_{sub}[s] \leftarrow H_{sub}[s] + p(D = d_j) * H(C \mid D = d_j)$

*3. Aggregate statistics across sub-samples*

$H \leftarrow mean(H_{sub}) + (2 \ln 2\ m)^{-1}$ # average taken over all sub-sample estimates

$\sigma_H \leftarrow std(H_{sub}) + (m/2)^{-1/2}$ # standard deviation across all sub-sample estimates